%% file: dpcmf.tex
\documentclass[10pt]{article}

\AtBeginDocument{%
  \providecommand\BibTeX{{%
    \normalfont B\kern-0.5em{\scshape i\kern-0.25em b}\kern-0.8em\TeX}}}

\usepackage{arxiv}
\usepackage{amsfonts}
\usepackage{amsmath}
\usepackage{multicol}
\usepackage{mathtools}
\usepackage{amsthm}
\usepackage{enumitem}
\usepackage[round]{natbib}

\usepackage{url}
\usepackage{hyperref}       
\hypersetup{
  colorlinks   = true, 
  urlcolor     = blue, 
  linkcolor    = blue, 
  citecolor   = blue 
}

\input{macros}

\theoremstyle{plain}
\newtheorem{theorem}{Theorem}[section]

\theoremstyle{definition}
\newtheorem{definition}[theorem]{Definition}

\theoremstyle{remark}
\newtheorem{remark}[theorem]{Remark}

\usepackage{microtype}
\usepackage{graphicx}
\usepackage{subcaption}
\usepackage{booktabs} 

\usepackage{algorithm}
\usepackage[algo2e,ruled,vlined,noend]{algorithm2e}

\title{Private Matrix Factorization with Public Item Features}

\author{
    Mihaela Curmei\thanks{UC Berkeley; Work done while at Google.} \hspace{0.5cm}
    Walid Krichene\thanks{Google.} \hspace{0.5cm}
    Li Zhang\thanks{Microsoft. Work done while at Google.} \hspace{0.5cm}
    Mukund Sundararajan\footnotemark[2]
}

\begin{document}
\maketitle
\begin{abstract}
We consider the problem of training private recommendation models with access to public item features.
Training with Differential Privacy (DP) offers strong privacy guarantees, at the expense of loss in recommendation quality. We show that incorporating public item features during training can help mitigate this loss in quality. We propose a general approach based on collective matrix factorization (CMF), that works by simultaneously factorizing two matrices: the user feedback matrix (representing sensitive data) and an item feature matrix that encodes publicly available (non-sensitive) item information.

The method is conceptually simple, easy to tune, and highly scalable. It can be applied to different types of public item data, including: (1) categorical item features; (2) item-item similarities learned from public sources; and (3) publicly available user feedback. Furthermore, these data modalities can be collectively utilized to fully leverage public data.

Evaluating our method on a standard DP recommendation benchmark, we find that using public item features significantly narrows the quality gap between private models and their non-private counterparts. As privacy constraints become more stringent, models rely more heavily on public side features for recommendation. This results in a smooth transition from collaborative filtering to item-based contextual recommendations.
\end{abstract}


.
\maketitle
\section{Introduction}

Recommender systems trained on private user feedback present the risk of leaking sensitive information about users' activity or preferences~\citep{zhang2021inference,calandrino2011privacy}, and thus, providing formal privacy protections is increasingly important. Differential privacy (DP)~\citep{dwork2014algorithmic} has emerged as the de facto standard for formalizing and quantifying privacy protections. These DP guarantees often come at the expense of some degradation in model quality, as DP training involves adding noise to quantities derived from user data (for example, adding noise to the gradients~\citep{abadi2016dpsgd}). Recent progress in private recommendation algorithms~\citep{jain2018differentially,chien2021private,krichene2023multi} has significantly improved the privacy/utility trade-offs, but there still remains a large quality gap between private models and their non-private counterparts.

It was recently shown~\citep{krichene2023multi, chien2021private} that these quality losses are due to degradation in item representation as a result of the noise added to ensure DP (particularly for tail items, which have fewer ratings and are more impacted by noise). Making item embeddings robust to the added noise may be the key to narrowing the quality gap between private and non-private models. One promising direction is to utilize public item features to improve item representation while maintaining strict user-privacy guarantees.

In this work, we investigate methods to utilize such item features to improve the quality of privacy-preserving recommenders. We take inspiration from the literature on Collective Matrix Factorization (CMF)~\citep{singh2008relational}, which learns shared embeddings from collections of related matrices, rather than a single matrix. Throughout the paper, we will distinguish between \emph{private user feedback}, which is sensitive and needs to be protected, and \emph{public item features}, which represent non-sensitive, publicly available information that does not need privacy protection.

\subsection{Contributions}

\begin{itemize}
\item \textbf{Formulation}:

We model both public item features and sensitive user-item feedback as matrices. Two low-rank factorizations are learned simultaneously. One factorization approximates the user feedback matrix and the other approximates the item feature matrix. Importantly, the item representation is shared between the two factorizations, which enables item embeddings to benefit from public features. This setup is versatile as it can encode various modalities of public information. For instance, features can represent public item metadata. The setup can also encode pairwise item similarity derived from public data, where the 'features' correspond to items and represent similarity scores. Finally, we can encode user feedback, for instance, from users who choose to make their ratings or reviews publicly available, here, the 'features' are users and represent the affinity between a user and an item.
\item \textbf{Method}: To provide DP guarantees, we propose Differentially Private Collective Matrix Factorization (DP-CMF), that extends the recently proposed DPALS algorithm~\citep{chien2021private} to the CMF formulation. DP-CMF works by adding noise to the sufficient statistics derived from sensitive data, while using exact statistics derived from public data.
\item \textbf{Evaluation}: We evaluate DP-CMF on the same private recommendation benchmark used in~\citep{jain2018differentially,chien2021private,krichene2023multi}. We find that incorporating public item features significantly narrows the quality gap between private and non-private models, particularly so when privacy requirements are high. This study offers a promising direction for improving privacy-utility trade-offs in recommender systems by leveraging public data sources while preserving user privacy.
\end{itemize}
 
\subsection{Related Work} 
\emph{Differential privacy in recommender systems.}
The importance of privacy in recommender systems has been recognized for a long time~\citep{narayanan2008netflix}, and some early attempts were made~\citep{mcsherry2009differentially,kapralov2013differentially} to develop differentially private algorithms that offer strong protection, but this usually required significant losses in model quality. Recent work~\citep{jain2018differentially,chien2021private,krichene2023multi} developed new algorithms that narrowed this quality gap, by using alternating minimization~\citep{chien2021private,jain2021private}, and developing methods to adaptively allocate privacy budgets~\citep{krichene2023multi}. Our proposed algorithm builds on these recent improvements, by extending the DPALS technique~\citep{chien2021private} to incorporate public item data. While utilizing public data to improve DP models has been explored in other domains (as described below), our work is the first to carry out a systematic study for private~recommenders.
\medskip

\noindent\emph{Using side features in recommenders.}
User and item side information are commonly employed to address the "cold-start" problem for users and items with limited or no interaction data \citep{gantner2010learning, saveski2014item, kula2015metadata, deldjoo2019movie, cortes2018cold}. Furthermore, side information can tackle fairness concerns and mitigate popularity bias in recommendations \citep{shi2014collaborative}. Side features can be integrated into MF models through Collective Matrix Factorization (CMF) \citep{singh2008relational, shi2014collaborative, dong2017hybrid, liang2016factorization, jenatton2012latent}, also known as Joint Matrix Factorization~\citep{zhu2007combining}, which originated in the Statistical Relational Learning literature~\citep{getoor2007introduction}. 
Our work leverages the CMF approach and extends it to private recommendations. While in recommender systems, both user and item side information can be useful, in the privacy context, it is more natural to consider only item side information, as it generally represents non-sensitive data, while user side information (such as demographic features) is sensitive and would require privacy protection. Our paper will hence focus on item features.

\medskip

\noindent\emph{Using public data to improve private models.}
Leveraging public information to enhance privacy/utility trade-offs has been explored in various contexts.
Existing approaches fall in two broad categories. The first is public pre-training followed by private fine-tuning. Empirically, this approach is effective in domains with abundant public data, such as natural language processing~\citep{li2021large, yu2021differentially, behnia2022ewtune} and vision~\citep{golatkar2022mixed, xu2022learning}. The second is to directly incorporate public data into the private learning process. These techniques are based either on projecting private gradients onto a low-dimensional subspace estimated from public gradients \citep{kairouz2020fast, yu2021not,zhou2021bypassing}, or utilizing public data to modify the objective function~\citep{bassily2020learning,amid2022public,li2022private}. For an extensive review, see~\citep{cummings2023challenges}. These approaches often make the restrictive assumption that public and private data come from the same distribution~\citep{kairouz2020fast,amid2022public,wang2020differentially,zhou2021bypassing} (so that public and private gradients lie on the same subspace). Our approach can work even if the public data comes from a different distribution: access to item metadata can be informative about item similarity, even if this data is of an entirely different nature than user feedback. Another notable difference is that existing work focuses on \emph{gradient-based} methods, while ours is, to the best of our knowledge, the first to explore the benefits of public data on \emph{second-order} methods (Alternating Least Squares).

\section{Preliminaries}\label{sec:setup}

\subsection{Setup \& Notation}
Throughout, $\boldM\in\mathbb{R}^{m\times n}$ denotes the user-item feedback matrix, and $\boldS\in\mathbb{R}^{s\times n}$ the item-feature matrix, where $m, n, s$ are the number of users, items, and features, respectively. We denote by $\Omega$ a subset of $[m] \times [n]$ representing the indices of the observed entries in $\boldM$. We define $\Omega_{i:} := \{j \in [m] : (i, j)\in \Omega\}$ the set of items rated by user $i$,and define $\Omega_{:j} := \{i \in [n] : (i, j)\in \Omega\}$ the set of users that rated item $j$. Further, we denote by $\Omega'$ the subset of $[s] \times [n]$ representing the observed entries of the item-feature matrix. For instance, $(k, j) \in \Omega'$ if item $j$ has corresponding public feature token $k$ (e.g., $j\equiv$ Titanic, $k\equiv$ director: James Cameron). The goal of CMF is to learn two low-rank factorizations: $\boldM_{\Omega} \approx \boldU \boldV^\top$ that approximates the user feedback matrix, and $\boldS_{\Omega'} \approx \boldF \boldV^\top$ that approximates the item feature matrix. Where $\boldU \in \mathbb{R}^{n\times d}$, $\boldV \in \mathbb{R}^{m\times d}$ and $F\in \mathbb{R}^{s\times d}$ are $d-$dimension embeddings corresponding to users, items and features, respectively. The notation $\boldM_{\Omega}$ means that approximate equality is desired only with respect to the entries $\boldM_{ij}$ for $(i,j)\in \Omega$.

For a vector $\boldv\in \mathbb{R}^d$, $\norm{\cdot}$ denotes the usual Euclidean $\ell^2$ norm. For two vectors $\boldu, \boldv \in \mathbb{R}^d$, $\ip{\boldu}{\boldv}$ and $\op{\boldu}{\boldv}$ denote the inner and the outer product, respectively. By $\Pi_{\text{PSD}}(\cdot)$, we denote the projection operation on the set of positive semidefinite matrices. By $\norm{\cdot}_{frob}$, we denote the Frobenius norm of a matrix. For a matrix $\boldU$, $\boldu_i$ specifies the $i$-th row. Finally, we use $\calN^d$ to denote the standard multivariate normal distribution and $\calN^{d\times d}$ to denote the distribution of symmetric $d\times d$ matrices whose upper triangular entries are i.i.d. standard normal.

\subsection{Privacy considerations}\label{sec:privacy}
Following~\citep{jain2018differentially,chien2021private,jain2021private}, we adopt the notion of \emph{user-level} DP~\citep{dwork2014algorithmic}, where the goal is to protect \emph{all of the ratings} from a user. Intuitively, the user-level DP guarantee limits the impact that any user can have on the algorithm's output. More formally, let $D = \{d_1, d_2, \ldots d_n\}$ be a set of inputs corresponding to the $n$ users, and let $\calA: \calD^n \to \calY$ be a randomized algorithm that produces an output $y \in \calY$. In our case, $d_i$ are the ratings associated with user $i$ and $y$ is the set of all item embeddings~$\boldV$ and feature embeddings~$\boldF$. Denote by $D_{-i}$ the inputs for all users except $i$. Two sets of inputs $D$, $D'$ are said to be \emph{adjacent} if they differ in at most one user; i.e. $D = \{d_i, D_{-i}\}$ and $D' = \{d_i', D_{-i}\}$.

\begin{definition}[User-level Differential Privacy \citep{kearns2014mechanism}]\label{def:jointDP}
An algorithm $\calA$ satisfies user-level $(\epsilon, \delta)$-DP if for all adjacent data sets $D$ and $D'$, and any measurable set of outputs $Y \subset \calY$, the following holds:
$\prob(\calA(D) \in Y) \leq e^{\epsilon}\prob(\calA(D') \in Y) + \delta.$
\end{definition}
Intuitively, $(\epsilon, \delta)$ are privacy parameters that control the "indistinguishability" between the outputs of the algorithm when it processes two datasets that differ in a single user's data. 
The smaller the values of $\epsilon$ and $\delta$, the stronger the privacy guarantee provided by the algorithm. The parameter $\delta$ is typically taken to be $\leq 1/n$ ($n$ is the number of users). The values of $\epsilon$ depend on the domain, studies typically report values ranging from $\epsilon = 0.1$ (high privacy regime) to $\epsilon=10$.\\

\begin{remark}[User-level vs. rating-level Differential Privacy]
Some prior techniques~\citep{dwork2014analyze,kapralov2013differentially} provide rating-level DP guarantees, meaning that neighboring datasets are allowed to differ in at most a single rating. In other words, rating-level DP limits risk of leakage from each individual rating, but this offers a much weaker protection at the user-level, (since users typically have many ratings, and the leakage risk compounds with the number of ratings). In contrast, user-level DP~\citep{kearns2014mechanism,jain2018differentially} ensures that a user's full set of ratings is protected. This makes user-level DP both more challenging to accomplish, but also more practically significant and relevant in terms of privacy protection of a user's data.
\end{remark}

\section{Differentially Private Collective Matrix Factorization}
We now introduce the DP-CMF algorithm for private recommendations with public item features. We first recall the Alternating Least Squares (ALS) algorithm for (non-private) CMF, then introduce the necessary modifications to satisfy user-level~DP.

\subsection{ALS for (non-private) CMF}\label{sec:als-cmf}

CMF jointly optimizes the following weighted loss function to find low-rank approximations of $\boldM_{\Omega}\approx \boldU \boldV^\top$ and $\boldS_{\Omega'}\approx \boldF \boldV^\top$.
\begin{equation}\label{eq:opt}
\begin{split}
\calL(\boldU,\boldV, \boldF) = & \sum\limits_{(i,j)\in \Omega}\boldW_{ij}\left(\ip{\boldu_i}{\boldv_j}-\boldM_{ij}\right)^2 + \alpha\sum\limits_{(k,j)\in \Omega'}\left(\ip{\boldf_k}{\boldv_j}-\boldS_{kj}\right)^2 +\\
&+\lambda \left( \norm{\boldU}^2_{frob} + \norm{\boldV}^2_{frob}\right) + \lambda'\norm{\boldF}^2_{frob},
\end{split}
\end{equation}
where $\boldW_{ij}$ is the weight associated with the contribution of user $i$'s rating of item $j$ to the loss function, $\lambda$ is the regularization weight for user and item embeddings and $\lambda'$ is the regularization weight for feature embeddings. Finally, $\alpha$ is a hyper-parameter that controls the relative importance of fitting the public versus private data. A small $\alpha$ means that item embeddings $\boldV$ will primarily depend on user-item feedback; whereas a large $\alpha$ means that item embeddings will depend more on the item-feature matrix. Although the loss is not jointly convex, for fixed item embeddings $\boldV$, it is a convex quadratic with respect to $(\boldU, \ \boldF)$ and vice-versa. ALS takes advantage of this fact, and alternates between updating $(\boldU, \boldF)$ and updating $\boldV$, as follows $\forall i \in [n], \forall k \in [s]$ and $\ \forall j \in [m]$, respectively :

{\small\begin{align}
\boldu_i^t  \xleftarrow{}& \arg\min_{u} \sum_{j\in\Omega_{i:}}\left(\boldW_{ij}\ip{\boldu}{\boldv_j^{t-1}}-\boldM_{ij}\right)^2 + \lambda\norm{\boldu}^2_2 \label{eq:cmf_updates_a} =  \Bigg[\sum_{j\in \Omega_{i:}}\boldW_{ij}\boldv_j^{t-1}\otimes\boldv_j^{t-1} + \lambda I \Bigg]^{-1}\Bigg[\sum_{j\in\Omega_i}\boldW_{ij}\boldM_{ij}\boldv_j^{t-1}\Bigg]; \\
\boldf_k^t  \xleftarrow{} &\arg\min_{f} \alpha  \sum_{j\in\Omega'_{k:}}\left(\ip{\boldf}{\boldv_j^{t-1}}-\boldS_{kj}\right)^2 + \lambda'\norm{\boldf}^2_2 \label{eq:cmf_updates_a2}= \Bigg[\sum_{j\in \Omega'_k}\boldv_j^{t-1}\otimes\boldv_j^{t-1} + \lambda' I \Bigg]^{-1}\Bigg[\sum_{j\in\Omega'_{k:}}\boldS_{kj}\boldv_j^{t-1}\Bigg];   \\
\boldv_j^t  \xleftarrow{} &\arg\min_{v} \sum_{i\in\Omega_{:j}}\left(\boldW_{ij}\ip{\boldu_i^{t}}{\boldv}-\boldM_{ij}\right)^2 +\alpha \sum_{k\in\Omega'_{:j}}\left(\ip{\boldf_k^t}{\boldv}-\boldS_{kj}\right)^2 + \lambda\norm{\boldv}^2_2\notag =\left[A^t_j\right]^{-1} \left[b^t_j\right];  \label{eq:cmf_updates_b}
\end{align}
where $ A^t_j := \Bigg[\sum_{i\in \Omega_{:j}}\boldW_{ij}\boldu_i^t\otimes \boldu^t_i + \alpha\sum_{k\in \Omega'_{:j}}\boldf^t_k\otimes\boldf_k + \lambda I \Bigg]^{-1}$ 
and $b^t_j :=  \Bigg[\sum_{i\in \Omega_{:j}} \boldW_{ij}\boldM_{ij} \boldu_i^t + \alpha\sum_{k\in \Omega'_{:j}}\boldS_{kj}\boldf^t_k\Bigg]$
}

\noindent The ALS updates for user and feature embeddings (Eqs.~ \eqref{eq:cmf_updates_a} and~\eqref{eq:cmf_updates_a2}) are decoupled and can happen simultaneously. In essence, item features (e.g. genre:comedy) can be treated as ``fictitious users''.\\

\begin{remark}[Implicit feedback and binary features]
When the user feedback is implicit (e.g. clicks, views), or when the public item features are categorical, we use the implicit ALS formulation~\citep{hu2008ials} that penalizes non-zero predictions outside of the observation sets $\Omega$ and $\Omega'$, by adding terms $\norm{\boldU \boldV^\top}_{frob}$ and $\norm{\boldF \boldV^\top}_{frob}$ to the optimization objective in Eq.~\eqref{eq:opt}. This results in changes to the update equations that are standard in the literature. 
\end{remark}

\begin{algorithm}[tb]
\SetAlgoLined
\DontPrintSemicolon
\SetKwProg{myproc}{Procedure}{}{}
\begin{flushleft}
{\bfseries Input:} User-item matrix $\boldM$; feature-item matrix $\boldS$; feature weight $\alpha$; weight matrix $\boldW$; initial item embeddings $\boldV^0$; number of steps $T$; clipping parameters $\Gamma_M$,$\Gamma_U$; regularization parameters $\lambda$,$\lambda'$.\\
\For{$t=1$ {\bfseries to} $T$}{
\nl Broadcast $\boldV^{t-1}$ to all users.
\\\nl Each user $i\in [n]$ updates (client-side):  \\
 $\boldu_i^t \xleftarrow{} \left[\sum_{j\in \Omega_i}\boldv_j^{t-1}\otimes\boldv_j^{t-1} + \lambda I \right]^{-1}\left[\sum_{j\in\Omega_i}\boldM_{ij}\boldv_j^{t-1}\right]$\label{step:user_update}
\\\nl Update feature embeddings (server-side): \\
$\boldf_k^t \xleftarrow{} \left[\sum_{j\in \Omega'_k}\boldv_j^{t-1}\otimes\boldv_j^{t-1} + \lambda' I \right]^{-1}\left[\sum_{j\in\Omega'_k}\boldS_{kj}\boldv_j^{t-1}\right]; k \in [s]$\label{step:feature_update}
\\\nl Update item embeddings (server-side): : \\
$\boldV^t \xleftarrow{}$ DPItemUpdate$(\boldM, \boldU^t, \boldF^t)$\label{step:item_update}
} 
\nl {\bfseries return } $\boldV^T, \boldF^T$\\
\medskip
\myproc{\emph{DPItemUpdate}$(\boldM, \boldU, \boldF)$}{
\nl Clip rating entries: $\boldM_{ij} = \min(\boldM_{ij}, \Gamma_M); \quad \forall i, j\Omega$\label{step:clip_rating}
\\\nl Clip user norms: $\boldu_i \xleftarrow{} \max(\norm{\boldu_i}, \Gamma_U)\frac{\boldu_i}{\norm{\boldu_i}}; \quad \forall i \in [n]$ \label{step:clip_user}
\\\For{$j=1$ {\bfseries to} $m$}{
\nl Sample Gaussian noise $\boldG_j\sim \Gamma_U^2\calN^{d\times d}, \quad \boldg_j \sim \Gamma_U\Gamma_M\calN^d$ \label{step:ssp_start}
\\\nl Compute noisy statistics: $\hat \boldA_{j}= \Pi_{\text{PSD}}\left[\sum_{i\in \Omega_j} \boldW_{ij} \boldu_i\otimes \boldu_i + \boldG_j + \lambda I\right]$,\ 
$\hat b_{j} = \sum_{i\in \Omega_j} \boldW_{ij} \boldM_{ij} \boldu_i + \boldg_j$ \label{step:noisy}
\\\nl Compute exact statistics: $\boldA_j' = \sum_{k\in \Omega'_j}\boldf_k\otimes\boldf_k + \lambda' I, \quad \boldb_j' =\sum_{k\in \Omega'_j}\boldS_{kj}\boldf_k$ \label{step:exact}
\\\nl Update item embedding: $\boldv_{j} \xleftarrow{} [\hat \boldA_j + \alpha \boldA_j']^{-1}[\hat \boldb_j + \alpha \boldb_j']$ \label{step:ssp_end}
} 
\nl {\bfseries return } $\boldV$
}
\end{flushleft}
\caption{Differentially Private CMF with Public Item Features}\label{alg:dp-cals}
\end{algorithm}

\subsection{Differentially Private CMF}\label{sec:dp-als}
To ensure user-level DP, we introduce DP-CMF (see Algorithm \ref{alg:dp-cals}), which extends the DPALS procedure~\citep{chien2021private} to CMF with public features. DP-CMF computes and releases the item and feature embeddings $(\boldV^t, \boldF^t$) with DP protection on a trusted centralized platform (server-side). Meanwhile, each user $i$ independently updates their embedding $\boldu_i^t$ on their own device (client-side). As a result, the user embedding update (step~\ref{step:user_update} of Algorithm \ref{alg:dp-cals}) is identical to the non-private update in Eq. \eqref{eq:cmf_updates_a} with additional assumption that the update is unweighted (i.e. $W_{ij}=1$). Furthermore, the feature embedding update (step~\ref{step:feature_update}) only depends on $\boldS$ (public data) and~$\boldV^{t-1}$ (which is DP-protected), hence by the DP post-processing property, it requires no additional noise and can be computed as in Eq.~\eqref{eq:cmf_updates_a2}.

On the other hand, the item embedding update (step~\ref{step:item_update}) depends on private data $\boldM, \boldU^t$, and must be modified to guarantee DP. This requires two modifications: the first is to limit the impact of each user on the item embeddings, this is done by clipping the magnitude of individual ratings (step~\ref{step:clip_rating}) clipping the user embedding norm (step~\ref{step:clip_user}) and weighting the ratings of each user with appropriately chosen weights $\boldW$ (step~\ref{step:noisy}). The second is to add noise to the sufficient statistics  (steps~\ref{step:ssp_start}-\ref{step:ssp_end}) via the Gaussian mechanism  \citep{vu2009differential, foulds2016theory, wang2018revisiting}. Note that the statistics $\hat \boldA_j, \hat \boldb_j$ (step~\ref{step:noisy}) depend on sensitive data and are protected via noise, while the statistics $\boldA'_j, \boldb'_j$ (step~\ref{step:exact}) depend only on public data and are computed exactly. \\

\begin{remark}
Step~\ref{step:ssp_end} intuitively highlights the potential benefit of using item features: the item embedding is the solution of a linear system $A \boldx=b$ with $A= \hat \boldA_j + \alpha \boldA'_j$, and $b=\hat \boldb_j + \alpha \boldb_j'$, where $\hat \boldA_j, \hat \boldb_j$ are noisy quantities derived from user feedback, while $\boldA_j', \boldb_j'$ are derived from public features and are exact. A larger $\alpha$ makes the solution more robust to the noise, but favors fitting the item features. When the item features are informative (e.g., they accurately capture item-item similarity), this can improve the item representation compared to only using noisy user feedback $(\alpha = 0)$.\\
\end{remark}

\begin{remark}[Computational cost of DP-CMF]
One step of DPALS\citep{chien2021private} consists of forming the sufficient statistics (a cost of $O(|\Omega|d^2)$) then solving $m+n$ linear systems (a cost of $O((m+n)d^3)$. In DP-CMF (Algorithm~\ref{alg:dp-cals}), the sufficient statistics computation cost increases to $O((|\Omega| + |\Omega'|d^2)$, and the linear system cost increases to $O((m+n+s)d^3)$. Hence, the added cost of using public features remains reasonable if $|\Omega'|$ is comparable in size to $|\Omega|$, and the total number of features $s$ is smaller or comparable to $m+n$.\\
\end{remark}

\begin{remark}[Threat model]
Observe that in this model, the recommendation platform broadcasts the item embeddings~$\boldV$ and the feature embeddings~$\boldF$. The user embeddings $\boldU$ are never published. Rather, each user $i$ can compute her own embedding $\boldu_i$ (by solving a least-squares problem involving her own ratings along with the published item embeddings $\boldV$, see Eq.~\eqref{eq:opt}), then use it to generate recommendations by computing scores $\boldu_i^\top \boldV$. This captures a very strong notion of privacy, as it protects user $i$ even against potential collusion of the remaining $n-1$ users (i.e. an adversary with access to $\boldV,\boldF$ and $D_{-i}$), while allowing the user to take full advantage of her data to generate recommendations. Importantly, the platform hosting the recommendation system is a trusted entity (it has access to the raw user ratings and user embeddings when computing the noisy sufficient statistics). The goal is to protect against privacy attacks from malicious users or external agents, not the recommender system itself. However, if the recommender itself is considered untrusted, these algorithms (DP-ALS and DP-CMF) can potentially be implemented using secure aggregation algorithms~\citep{bonawitz2017}, although this comes at an increased computational cost.\\
\end{remark}

\begin{prop}[Privacy Guarantee]
For all $\epsilon > 0, \delta \in (0, 1)$, if the inputs to Algorithm~\ref{alg:dp-cals} satisfy $\sum_{j \in \Omega_i} \boldW_{ij}^2 \leq \frac{\epsilon^2}{4T(\log(1/\delta)+\epsilon)} \forall i \in [n]$, then the algorithm is $(\epsilon, \delta)$ user-level DP.
\end{prop}

\begin{remark}
The weights $\boldW$ are used to control a user's impact on the model. The simplest way to generate weights that satisfy the condition of the proposition is to assign a uniform weight to each user. Specifically, given a desired privacy level $(\epsilon, \delta)$, let $\beta = \frac{\epsilon^2}{4T(\log(1/\delta) + \epsilon)}$, then simply set $\boldW_{ij} = \sqrt{\beta / |\Omega_i|}$ satisfies the inequality (notice that a user with more ratings, i.e. a larger $|\Omega_i|$, will have lower weights, to limit the solution's sensitivity to that user's data). A more sophisticated method was developed in~\citep{krichene2023multi} that adapts to the item frequencies by putting more weight on infrequent items. We use the latter in our experiments.
\end{remark}

\begin{proof}

First, we argue that it suffices to prove the result for $\alpha = 0$. Indeed, by step~\ref{step:exact}, the statistics $\boldA_j', \boldb_j'$ only depend on the public feature matrix $\boldS$ and on the feature embeddings $\boldF^{t}$, which in turn only depends on $\boldS$ and $\boldV^{t-1}$ (by step~\ref{step:feature_update}). Since $\boldV^{t-1}$ is released with DP protection, there is no additional privacy cost for computing $\boldA_j', \boldb_j'$ (by the post-processing property of DP~\cite[Proposition 2.1]{dwork2014algorithmic}). Therefore the privacy guarantee of the algorithm with $\alpha = 0$ or $\alpha > 0$ are identical. When $\alpha = 0$ (no features), the algorithm becomes identical to DPALS, and the guarantee is proved in~\cite[Theorem 3.3]{krichene2023multi}.
\end{proof}

\section{Empirical Evaluation}\label{sec:exp}

We evaluate DP-CMF on a standard DP recommendations benchmark used in~\citep{jain2018differentially,chien2021private,krichene2023multi}, based on the MovieLens datasets \citep{harper2015movielens}. The benchmark considers a rating prediction task on the MovieLens 10M (\ml{10}) dataset, which records over $10$ million ratings ranging from 1 to 5 for $n=69878$ users and $m=10677$ movies. For the feature-item matrix  we consider 3 sources of public data: \\

\emph{Item metadata.}
We construct a categorical feature dataset by cross-referencing movie IMDb identifiers with data available on \href{https://www.wikidata.org/}{Wikidata.org}. For each movie in the \ml{10} dataset, we collect genre, topic, and cast information. We construct a binary feature matrix $\boldS$, where each row corresponds to a feature token (e.g., the first row is labeled as \texttt{director:James Cameron}, and non-zero entries in this row correspond to movies directed by James Cameron). The metadata dataset comprises $s=12637$ feature tokens with an overall feature density of $0.13\%$.\\
 
\emph{Item to item similarity scores.}
We create an item-to-item similarity dataset from a non-private recommendation model trained on a variant of the \ml{20} dataset, as proposed in \citep{liang2018variational}. This dataset, is commonly used for benchmarking recommender performance on implicit feedback, as the training data is a binary matrix corresponding to ratings $\geq 4$. We first train a Matrix Factorization model on the dataset and use the resulting item embeddings to identify, for each movie, the $k$ most similar movies based on similarity scores (we experiment with inner product and cosine similarity). Each row in the feature matrix $\boldS$ corresponds to a movie, with non-zero values $\boldS_{ij}$ indicating similarity between movies $i$ and $j$. Finally, we consider both actual and binarized scores in the $\boldS$ feature matrix.\\
 
\emph{Public user ratings.}
We use the \ml{20} dataset and select the observations corresponding to 70152 users that do not overlap with the \ml{10} users. We consider this set of user-item feedback as public item side data. More specifically, each user, whose data is considered public, plays the same functional role as a feature token. We observe that in this case, the public data is of the same semantic type as the private data. This setup is closest to the common assumption in the literature, that private and public data come from the same or similar distributions. In experiments, we use subsets of public users of various sizes, ranging from very small values ($s=100$) to the full available data ($s=70152$). Finally, we consider both raw ratings (from the original \ml{20}) as well as binarized ratings. 

\subsection{Experimental procedure} 
We follow the procedure of \citep{lee2016llorma} to partition the \ml{10} into train, test and validation datasets. For the privacy parameters, following~\citep{jain2018differentially,chien2021private,krichene2023multi}, we consider a range of $\epsilon = [1,5, 10, 20]$ and fix $\delta= 10^{-5}$. For each privacy setting we use the optimal hyper-parameters such as number of ALS iterations $T$, regularization weight $\lambda$, clipping norm $\Gamma_U$ tuned by \citep{krichene2023multi} for the same task without side features. With these pre-tuned hyper-parameters, for each $\epsilon$ and each public data source, we tune only\footnote{Retuning the full set of hyper-parameters may lead to even stronger performance, but our experiments show that tuning only the two parameters $\alpha, \lambda'$ already achieves quality improvements. This leads to a simple procedure, where one can first tune the DP-ALS hyper-parameters, then separately tune the CMF-related parameters $\alpha, \lambda'$.} the hyper-parameters corresponding to side features: the side feature weight $\alpha$ and side feature regularization $\lambda'$. We select hyper-parameters based on performance on validation set and finally we report performance on test set. Performance is measured using the Root Mean Squared Error (RMSE) between true and predicted ratings: $RMSE(\boldU, \boldV) = \sqrt{\frac{1}{\vert \Omega\vert}\sum_{i,j \in \Omega}(\boldM_{ij} - \ip{\boldu_i}{\boldv_j})^2}$. It is important to note that, although the training loss considers side feature-item data and the learned embeddings for features play a crucial role in updating item embeddings, the final performance is measured solely based on rating data (user feedback). In addition, we report performance metrics sliced by item popularity to gain a deeper understanding how public item information impacts the the quality of private models across frequent and infrequent items.

\subsection{Results}
\begin{figure}
    \centering
    \begin{subfigure}[t]{0.49\linewidth}
    \centering
    \includegraphics[width = .96\linewidth]{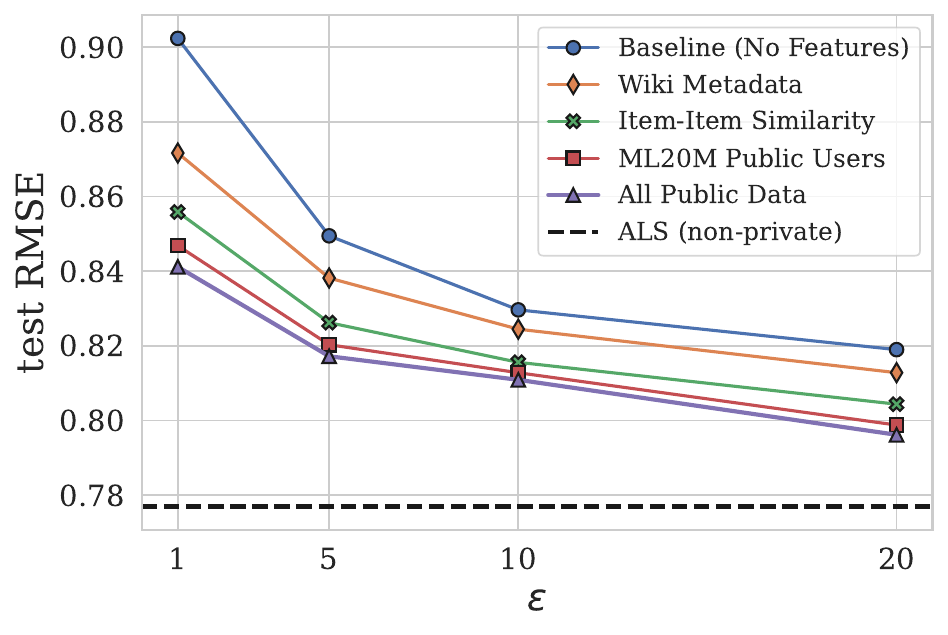}
    \caption{Privacy-accuracy trade-offs on \ml{10}}
     \label{fig:cmf_perf_a}
 \end{subfigure}
    \begin{subfigure}[t]{0.49\linewidth}
    \centering
  \includegraphics[width = .96\linewidth]{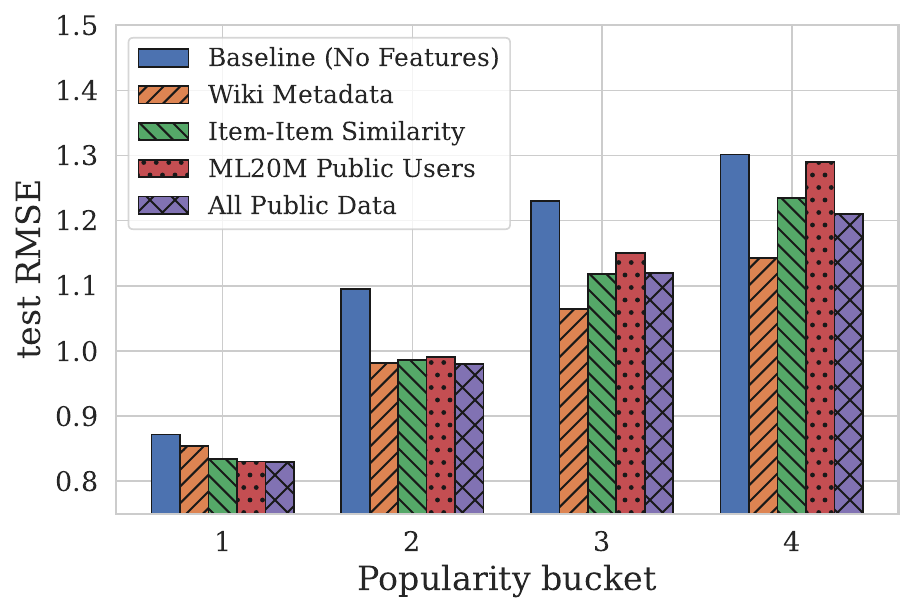}
    \caption{Model performance for $\epsilon=1$ by item popularity}
    \label{fig:cmf_perf_b}
    \end{subfigure}
    \caption{Impact of public item features on private recommendation accuracy.  Wiki Metadata corresponds to categorical genre, topic and cast features. Item-Item Similarity considers $k=100$ similar items according to dot product; ML20M Public Users considers binary observations for ratings $\geq4$}
    \label{fig:cmf_perf}
\end{figure}
In Figure \ref{fig:cmf_perf_a}, we compare DP-CMF's performance to the DPALS algorithm without side features from~\citep{chien2021private, krichene2023multi}  (blue curve), which is the current state-of-the-art on the \ml{10} benchmark. We also report for reference the non-private ALS baseline (dashed line). Incorporating public item information significantly narrows the existing quality gap between private and non-private models. The relative improvement depends on the public data source. Public user rating data (red curve) consistently outperforms other sources, as expected, since it comes from a closely related distribution. However, even tangentially related public item data, such as item metadata from Wikidata, substantially improves model quality. Furthermore, the different public item data modalities are composable, leading to compunded accuracy improvements (purple curve). The gap between private and non-private models is largest for high privacy requirements (low $\epsilon$), with side features closing up to 60\% of the performance gap. 
\begin{figure}
    \centering
    \begin{subfigure}[t]{0.4\linewidth}
    \centering
    \includegraphics[width=0.78\linewidth]{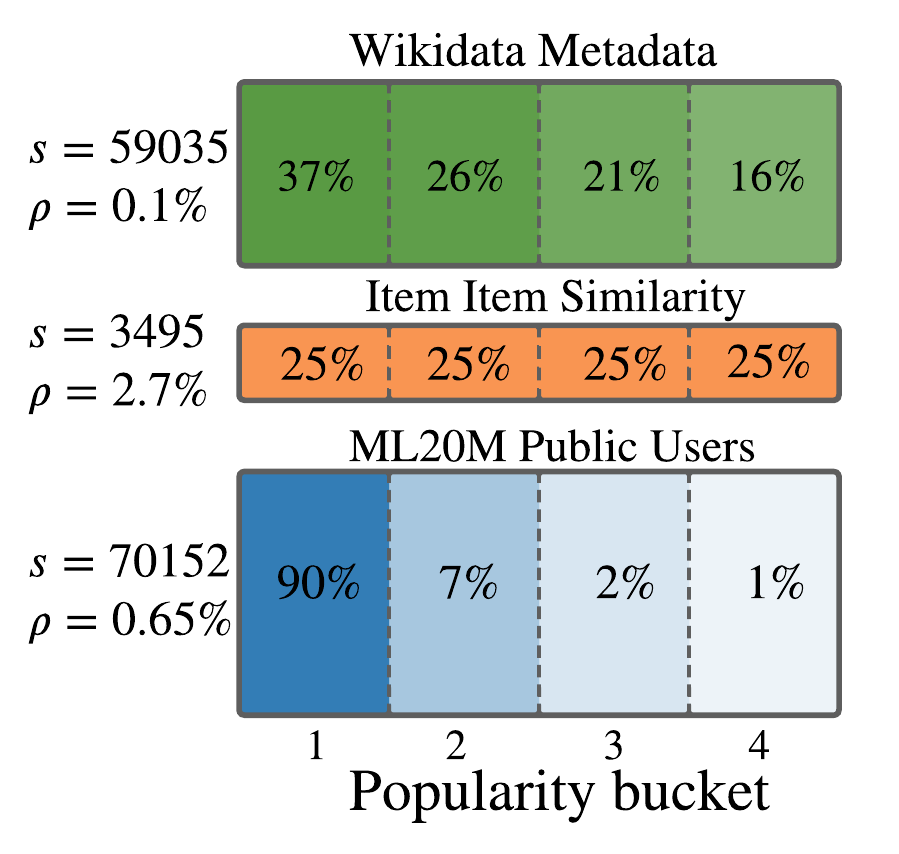}
    \caption{Feature matrix density across popularity levels.}
    \label{fig:f_density}
    \end{subfigure}
     \hspace{0.2cm}
    \begin{subfigure}[t]{0.45\linewidth}
    \centering
    \includegraphics[width=0.78\linewidth]{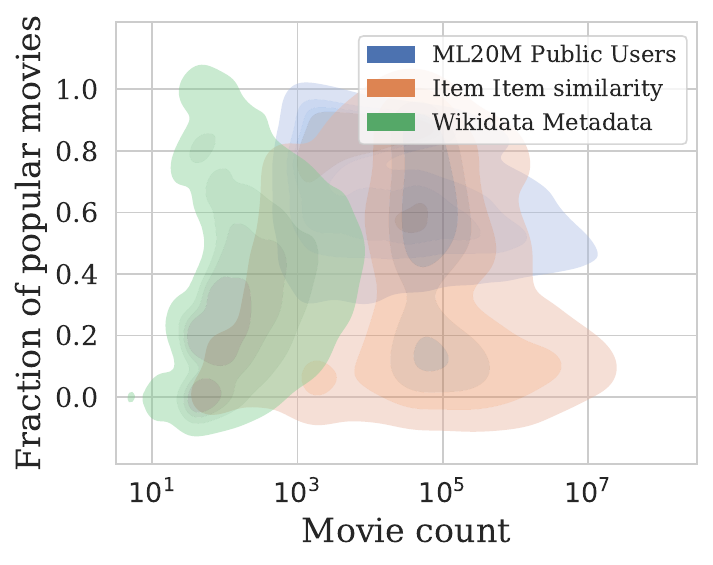}
    \caption{Share of popular items vs. feature prevalence.}
    \label{fig:f_frac_pop}
    \end{subfigure}
\caption{Popularity bias of Wikidata metadata, item-item similarity and \ml{20} public users' data}
\label{fig:f_bias}
\end{figure}

\subsubsection{Tail performance}

Figure \ref{fig:cmf_perf_b} shows performance across four popularity buckets for models trained under privacy parameter $\epsilon=1$, with each bucket containing roughly 2500 items. Due to the skewed distribution of ratings, the buckets hold 86.6\%, 9.4\%, 3\%, and 1\% of the all ratings, respectively. Thus, popular items have a greater impact on overall performance. The performance ordering of head items (bucket 1) matches the global ordering. However, for tail items (buckets 2 through 4), the order is reversed, with Wikidata features showing the most improvement for tail items. 

We posit that Wikidata movie metadata outperforms on tail items due to its less pronounced bias towards popular items. As Fig. \ref{fig:f_density} illustrates, 90\% of feature-item observations from \ml{20} public users correspond to popular items, while only 37\% of Wikidata feature matrix entries do so. However, feature density alone doesn't fully account for this performance difference, as item-to-item similarity doesn't perform as well on tail items, despite being perfectly balanced (by construction, we select the same number of neighbors for all movies).One hypothesis is that side features are most beneficial for tail items when they can transfer information from popular items. Fig. \ref{fig:f_frac_pop} supports this, showing that while public users primarily rate popular movies. Wikidata features, on the other hand, are less frequent, but describe both popular and tail items. This balance may explain why Wikidata outperforms other public data sources on tail items.
Another notable difference between the sources of data is illustrated in Fig. \ref{fig:f_frac_pop}. The figure shows, for each feature, the fraction of top-bucket movies for that feature (so a fraction of $r$ means that among all occurrences of that feature, $r$ fraction fall in the top bucket while the rest fall in other buckets). We can observe that while public users primarily rate popular movies, Wikidata features are more balanced across popular and tail items. This may explain why Wikidata outperforms other data sources on tail items.

\subsubsection{Performance across public data sources}
\begin{figure}
    \centering
    \begin{subfigure}[t]{0.45\linewidth}
    \centering
    \includegraphics[width=0.77\linewidth]{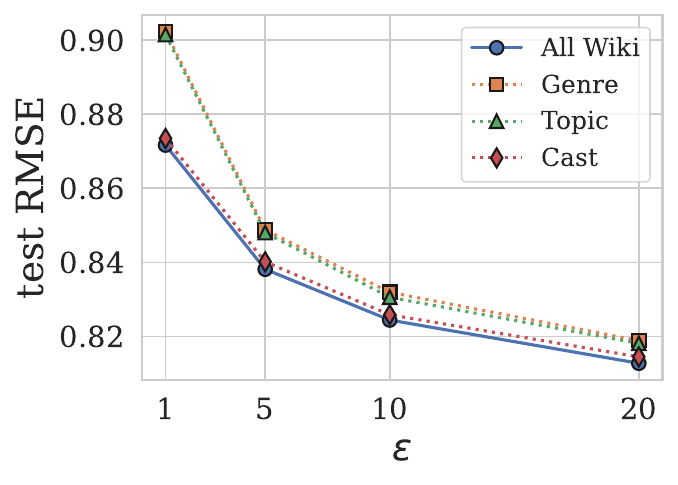}
    \caption{Privacy-Accuracy Tradeoffs}
    \label{fig:wiki_by_eps}
    \end{subfigure}
    \begin{subfigure}[t]{0.45\linewidth}
    \centering
    \includegraphics[width=0.77\linewidth]{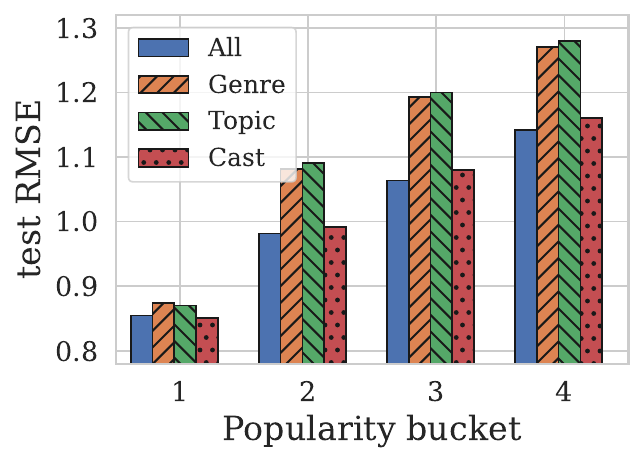}
    \caption{Sliced RMSE ($\epsilon=1$)}
    \label{fig:wiki_sliced}
    \end{subfigure}
\caption{Performance comparison of DP-CMF using Genre, Topic, and Cast data from Wikidata Metadata}
\label{fig:wikidata}
\end{figure}

We find that cast information alone captures most of the performance lift achieved by Wiki Metadata features (\ref{fig:wikidata}). The cast information is the most effective side feature for both head and tail items. This phenomenon is potentially explained by the fact that cast information is very granular and plausibly correlates with user preferences. 

\begin{figure}
    \centering
    \begin{subfigure}[t]{0.45\linewidth}
    \centering
    \includegraphics[width=0.77\linewidth]{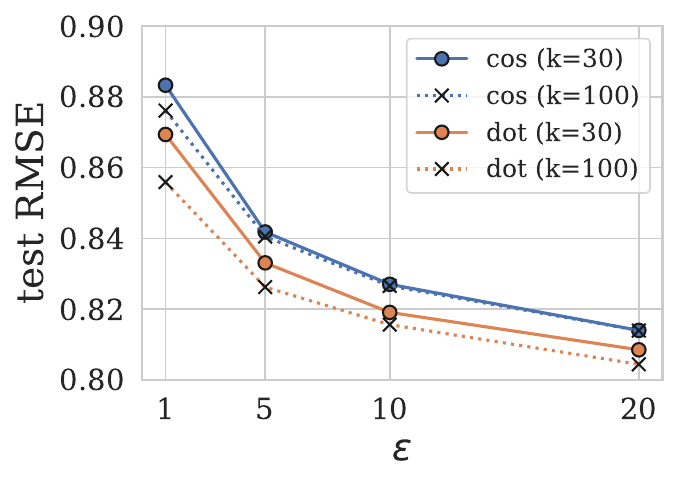}
    \caption{Privacy-Accuracy Tradeoffs}
    \label{fig:sim_by_eps}
    \end{subfigure}
    \begin{subfigure}[t]{0.45\linewidth}
    \centering
    \includegraphics[width=0.77\linewidth]{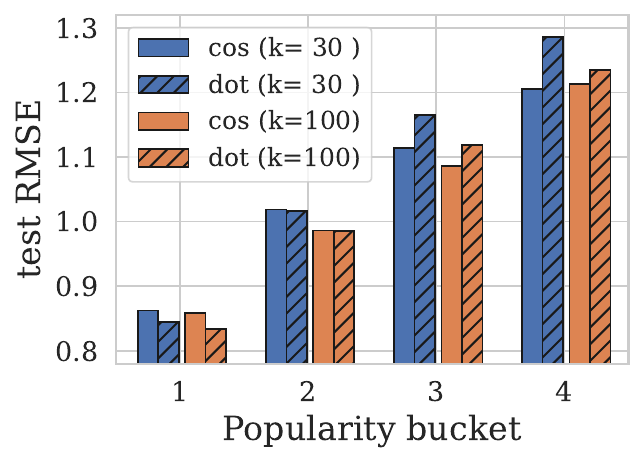}
    \caption{Sliced RMSE ($\epsilon=1$)}
    \label{fig:sim_sliced}
    \end{subfigure}
\caption{Comparing DP-CMF performance across varying similarity functions and numbers of selected similar items}
\label{fig:itemsim}
\end{figure}

In Figure \ref{fig:itemsim} we consider variants of pairwise item similarity. We find that the performance improves with the number of similar items for both cosine and inner product similarity scoring. Inner product scores generally outperform cosine similarity scores, in part because they take into account the magnitude of two vectors, not just their angle. Given that higher magnitudes typically correspond to more popular items, this leads to  popularity bias, reflected in the comparatively weaker performance of dot product similarities on tail items.

\begin{figure}
    \centering
    \begin{subfigure}[t]{0.45\linewidth}
    \centering
    \includegraphics[width=0.77\linewidth]{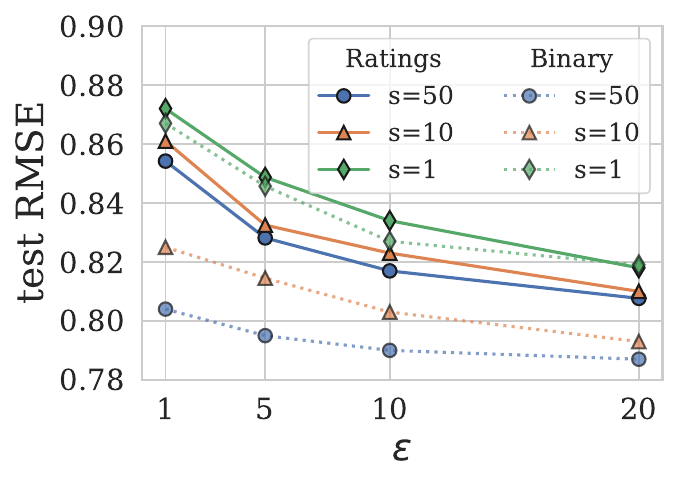}
    \caption{Privacy-Accuracy Tradeoffs}
    \label{fig:user_by_exp}
    \end{subfigure}
    \begin{subfigure}[t]{0.45\linewidth}
    \centering
    \includegraphics[width=0.77\linewidth]{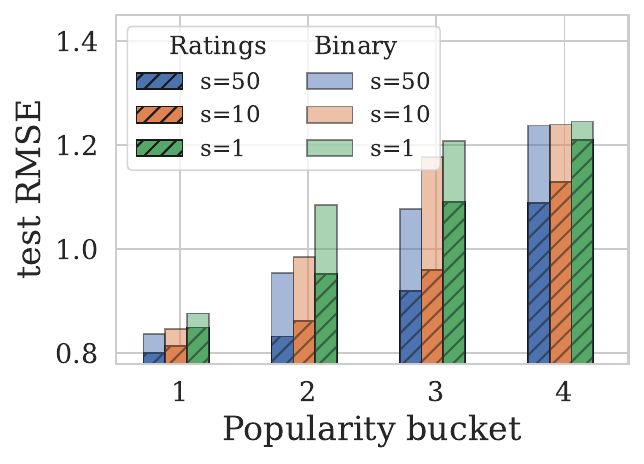}
    \caption{Sliced RMSE ($\epsilon=1$)}
    \label{fig:user_sliced}
    \end{subfigure}
\caption{Comparing DP-CMF performance for varying number of \ml{20} Public Users (in thousands)}
\label{fig:users}
\end{figure}

Finally, in Figure \ref{fig:users} we consider public item data derived from public ratings. Increasing the number of features (in this case, users with public ratings) significantly enhances model performance. This improvement is more pronounced when using non-binarized ratings, with the private model's performance approaching that of the non-private model even under strict privacy settings. While the most considerable accuracy gains are achieved with large amounts of public data ($s=50000$), even modestly sized sources of in-distribution data ($s=1000$) yield performance improvements comparable to the best gains achieved through Wiki Metadata.

\section{Discussion}
In this work, we introduce DP-CMF, a method aimed at improving the privacy-accuracy trade-off of private recommendation models. Our technique incorporates public item feature data into private recommendations that satisfy $(\epsilon, \delta)$-DP. This approach is simple to implement, easy to tune, and highly scalable.
DP-CMF allows for the integration of public side item information, pairwise item similarities, and public rating data. This is achieved within the same formulation, without requiring any changes in privacy accounting. Our experimental results demonstrate practical improvements in the privacy-accuracy trade-off by utilizing public item features.

Identifying public features that align with user interests and enhance recommendation performance remains a challenge. This task is domain-dependent. In general, access to high-quality annotations is beneficial, and this may be harder to obtain in some domains, for instance when content creation is cheap, and annotations are relatively more expensive. In such cases, another potential source is learning unsupervised, content-based similarity~\citep{jansen2018}

Future work includes comparing DP-CMF with pre-training approaches and extending our methodology beyond CMF. This could involve exploring other models that utilize item side features, such as Inductive Matrix Completion ~\citep{gantner2010learning,xu2013speedup,chiang2015matrix, jain2013provable,goldberg2010transduction} which enjoys favorable theoretical guarantees.

\pagebreak

\bibliographystyle{plainnat}
\bibliography{dpcmf}

\end{document}

%% file: macros.tex
\newcommand{\boldb}{\ensuremath{\boldsymbol{b}}}

\newcommand{\boldf}{\ensuremath{\boldsymbol{f}}}
\newcommand{\boldg}{\ensuremath{\boldsymbol{g}}}

\newcommand{\boldu}{\ensuremath{\boldsymbol{u}}}
\newcommand{\boldv}{\ensuremath{\boldsymbol{v}}}

\newcommand{\boldx}{\bfx}

\newcommand{\boldA}{\ensuremath{\boldsymbol{A}}}

\newcommand{\boldF}{\ensuremath{\boldsymbol{F}}}
\newcommand{\boldG}{\ensuremath{\boldsymbol{G}}}

\newcommand{\boldM}{\ensuremath{\boldsymbol{M}}}

\newcommand{\boldS}{\ensuremath{\boldsymbol{S}}}

\newcommand{\boldU}{\ensuremath{\boldsymbol{U}}}
\newcommand{\boldV}{\ensuremath{\boldsymbol{V}}}
\newcommand{\boldW}{\ensuremath{\boldsymbol{W}}}

\newcommand{\bfx}{\ensuremath{\mathbf{x}}}

\newcommand{\calA}{\ensuremath{\mathcal{A}}}

\newcommand{\calD}{\ensuremath{\mathcal{D}}}

\newcommand{\calL}{\ensuremath{\mathcal{L}}}

\newcommand{\calN}{\ensuremath{\mathcal{N}}}

\newcommand{\calY}{\ensuremath{\mathcal{Y}}}

\newcommand{\prob}{\operatorname{Pr}}

\newtheorem{lem}{Lemma}[section]
\newtheorem{prop}[lem]{Proposition}

\makeatletter
\newcommand{\vast}{\bBigg@{4}}
\newcommand{\Vast}{\bBigg@{5}}
\makeatother

\newcommand{\ip}[2]{\left\langle #1, #2\right\rangle}
\newcommand{\op}[2]{#1 \otimes #2}

\newcommand{\norm}[1]{\| #1 \|}

\renewcommand{\epsilon}{\varepsilon}

\newcommand{\ml}[1]{\texttt{ML{#1}M}}

